\documentclass%[twocolumn]
{amsart}
\usepackage{amssymb,amsmath,graphicx}
\theoremstyle{definition}

\theoremstyle{remark}

\numberwithin{equation}{section}
\begin{document}
%\enlargethispage{1in}
%\hspace*{115pt}\scalebox{.7}{\includegraphics{figures/Figure7}}
%\end{document}
\title [Superunification]{Outline of a Superunification Model}
\author{J. Towe}
%\\
\address{Department of Physics, The Antelope Valley College, Lancaster, CA 93536}%
\email{jtowe@avc.edu}\
%\thanks{}%
%\subjclass{}%
%\keywords{}%
%\date{}%
%\dedicatory{}%
%\commby{}%
\begin{abstract}

Tensor products of SM bosons and fermions form a spin-2
self-realization of SU(5),
%provided that one adopts three fermionic generations and
%preservation of generation by gravitational GUT interactions as
%necessary conditions. In this context moreover, SUGRA GUT
%interactions can be introduced at SM scale
and because every quark can be interpreted as a lepton that has
coupled to an appropriate element of this adjoint representation,
and inversely, SUGRA GUT interactions between proposed spin-2
elements and baryons of spin-(3/2) can exist as residual
manifestations of quark-lepton transitions that occur within
baryonic domains of asymptotic freedom. Such interactions preserve
baryonic structure and continually re-establish a locally SUSY
version of broken SU(5), indicating a recurring inflation event
that addresses the large-scale.

\end{abstract}\maketitle
%\\[-116 pt]
\section {SUGRA GUT Hypothesis}\label{S:intro}
%The three known generations of quarks and leptons form an
%irreducible representation of the Lorentz group:

%\[
%   \mathbf{A}=
%   \begin{bmatrix}
%      e^{-}_{L}D_{L}&&&&&&&&\\
%      e^{+}_{R}\overline{D}_{R}&0&0&0&0&0&0&0&0\\
%      &e^{-}_{R}D_{L}&&&&&&&\\
%      0&e^{+}_{L}\overline{D}_{R}&0&0&0&0&0&0&0\\
%      &&\mu^{-}_{L}S_{L}&&&&&&\\
%      0&0&\mu^{+}_{R}\overline{S}_{R}&0&0&0&0&0&0\\
%      &&&\mu^{-}_{R}S_{R}&&&&&\\
%      0&0&0&\mu^{+}_{L}\overline{S}_{L}&0&0&0&0&0\\
%      &&&&\tau^{-}_{L}B_{L}&&&&\\
%      0&0&0&0&\tau^{+}_{R}\overline{B}_{R}&0&0&0&0\\
%      &&&&&\tau^{-}_{R}B_{R}&&&\\
%      0&0&0&0&0&\tau^{+}_{L}\overline{B}_{L}&0&0&0\\
%      &&&&&&\nu^{e^{-}}_{L}U_{L}&&\\
%      0&0&0&0&0&0&\nu^{e^{+}}_{R}\overline{U}_{R}&0&0\\
%      &&&&&&&\nu^{\mu^{-}}_{L}C_{L}&\\
%      0&0&0&0&0&0&0&\nu^{\mu^{+}}_{R}\overline{C}_{R}&0\\
%      &&&&&&&&\nu^{\tau^{-}}_{L}T_{L}\\
%      0&0&0&0&0&0&0&0&\nu^{\tau^{+}}_{R}\overline{T}_{R}
%   \end{bmatrix}
%\]
Einstein's commitment to the a priori was reflected in many
comments; e.g. "... but the creative principle lies with
mathematics. In a certain sense therefore, I hold it true that
pure thought can grasp reality as the ancients dreamed." This
philosophy is not antithetical to the current aspirations of
physics. General relativity is the Noether principle that
preservation of GL(4) by the equations of physics is equivalent to
the conservation of 4-momentum by classical gravity and the
elusive SUGRA GUT theory is probably a Noether principle which
identifies a spin-2 adjoint representation of SU(5) with all
tensor products of standard model bosons and fermions that produce
spin-2 composites. Specifically if three generations of fermions
are incorporated, and if generation is preserved by gravitational
GUT interactions, then there are $5^{2}-1$ such composites:
\begin{equation} \label{E:int}
\sum_{\mu,\nu=0}^{3}T^{A}_{\alpha
B}g_{\mu\nu}^{\alpha}dx^{\mu}\otimes dx^{\nu}: A, B=1,...,5;
\alpha = 1,...,5^{2}-1,
\end{equation}
To demonstrate that this adjoint representation is more than a
formal result, it is shown that the above elements are components
of couplings that form a locally super-symmetric, adjoint
representation of SU(5). To establish this, it is observed that
every quark can be interpreted as a lepton that has coupled with
an appropriate spin-2 composite from the proposed adjoint
representation, and inversely; and that interactions between
spin-2 fields and spin-(3/2) baryons can therefore be interpreted
as residual manifestations of interactions between spin-2 fields
and quarks within baryonic domains of asymptotic freedom; i.e.
manifestations of quark-lepton transitions within baryonic domains
of asymptotic freedom. It can be demonstrated that there are
$5^{2}-1$ classes of SUGRA couplings (the number required for a
SUGRA self-realization of SU(5)) if and only if interactions
between spin-2 composites and spin-(3/2) baryons are restricted to
those that correspond to productions of leptons from quarks and
inversely. Baryons of spin-(1/2) can also be included in the
discussion provided that a photon is absorbed and radiated
together with each spin-2 anti-field (to preserve local
super-symmetry).

\par
It will now be demonstrated that SUGRA GUT interactions within
baryonic domains of asymptotic freedom produce quark-lepton
transitions that preserve baryonic structure, provided that the
four diagonal (preserved) generators of SU(5) are identified with
quantum numbers $I_{3}$, Y (strong hypercharge), F (fermion
number) and $\Delta$Q, where $\Delta$Q describes the difference
between the charges of a quark and lepton that share the same
values of the other three quantum numbers.

\section {Sub-Baryonic Interactions}\label{S:intro}

The eight composites that correspond to the light fermionic
generation are as follows:
\begin{equation} \label{E:int}
\gamma\otimes\gamma \otimes e^{-}_{L}\otimes \overline{D}_{R}
\end{equation}

\begin{equation} \label{E:int}
\gamma \otimes \gamma \otimes D_{L}\otimes e^{+}_{R}
\end{equation}

\begin{equation} \label{E:int}
\gamma\otimes\gamma \otimes e^{-}_{R}\otimes \overline{D}_{L}
\end{equation}

\begin{equation} \label{E:int}
\gamma\otimes\gamma \otimes D_{R}\otimes e^{+}_{L}
\end{equation}

\begin{equation} \label{E:int}
\gamma\otimes\gamma \otimes \nu^{e^{-}}_{L}\otimes
\overline{U}_{R}
\end{equation}

\begin{equation} \label{E:int}
\gamma\otimes\gamma \otimes U_{L}\otimes \nu^{e^{+}}_{R}
\end{equation}

\begin{equation} \label{E:int}
\gamma\otimes\gamma \otimes \nu^{e^{-}}_{R}\otimes
\overline{U}_{L}
\end{equation}

\begin{equation} \label{E:int}
\gamma\otimes\gamma \otimes U_{R}\otimes \nu^{e^{+}}_{L}.
\end{equation}
\par

The conservation laws that are introduced above permit three types
of interactions. These three types, which will be designated Type
A, Type B and Type C respectively preserve $I_{3}$=-1/2,
$I_{3}$=+1/2 and $I_{3}$=0. A Type A interaction is exemplified by
a locally SUSY vertex
\begin{equation} \label{E:int}
U_{L}\otimes D_{L}\otimes D_{L}+\gamma_{L}\otimes
\gamma_{L}\otimes\overline{U}_{R}\otimes
\nu^{e^{-}}_{L}\rightarrow \nu^{e^{-}}_{L}\otimes D_{L}\otimes
D_{L}+\gamma_{L}\otimes\gamma_{L},
\end{equation}
and a complementary vertex
\begin{equation} \label{E:int}
\nu^{e^{-}}_{L}\otimes D_{L}\otimes
D_{L}+\gamma_{L}\otimes\gamma_{L}\otimes
U_{L}\otimes\nu^{e^{+}}_{R}\rightarrow U_{L}\otimes D_{L}\otimes
D_{L}+\gamma_{L}\otimes \gamma_{L}.
\end{equation}
Clearly such interactions preserve baryonic structure, but do
these interactions predict anything that could subject the
proposed model to confirmation or disconfirmation? Possibly.
Typically, each baryon consists of many triplets. Until now it was
assumed that all triplets within a given baryon consist of the
same three quark flavors. But the proposed model seems to indicate
that at any given instant, some quarks may be replaced by
dynamical configurations of leptons and spin-2 composites. Thus
the proposed model seems to predict a fine-grained inhomogeneity
of density for the interiors of baryons. In this context, the
proposed theory may lend itself to confirmation.

\section {The Standard Model in Terms of Type IIB Strings}\label{S:intro}

\par
AdS/CFT correspondence [E. D'Hoker, 2002] admits a model of the
early universe in which a de Sitter sphere is equivalent to an
event horizon that encloses a world volume of coincident D3-branes
(a Type IIB string theory). If these D3 branes consist of three
color branes, two left branes, a right brane and a leptonic brane,
then the fermions of the standard model can be constructed by
intersections of these branes with open strings.  Due to the D3
nature of postulated branes moreover, it is argued that this
structure permeates 4-spacetime. A 6-dimensional space is
transverse to the world volume (this context parallels that
described by Zweibach and others). The corresponding CFT model is
in terms of a symmetry U(7), which is defined on the postulated
seven D3-branes and implicitly then, on all of 4-spacetime. It is
argued that Planck scale was introduced a priori onto 4-spacetime
in the limit where radii of curvature of postulated 4-dimensional
hyper-spheres (Section 5) are very small [J. Towe, 2006].
\par
Given a string background, every super Yang-Mills theory
corresponds to a Kaluza-Klein theory on a (4+k)-dimensional
product space, where k is the dimension of the space that is
transverse to the event horizon and to the enclosed world volume.
It is instructive to consider the Dirac operator
\begin{equation} \label{E:int}
i\Gamma^{A}D_{A}=i\gamma^{\mu}D_{\mu}(x)+i\gamma^{m}D_{m}(y),
\end{equation}
which is defined on the (4+6)-dimensional manifold of the
Kaluza-Klein theory that is relevant to the present discussion:
\begin{equation} \label{E:int}
\sqrt{detg_{AB}}R_{AB}g^{AB}=\sqrt{detg_{\mu\nu}det\gamma_{mn}}[R_{4}(x)+R_{N}(y)+higher
order terms].
\end{equation}
Specifically, it is instructive to consider the eigenvalues of the
Dirac operator on the 6-dimensional manifold that is transverse to
the event horizon. In Planckian units, these eigenvalues are
\begin{equation} \label{E:int}
i\Gamma^{A}D_{A}=i\gamma ^{\mu}D_{\mu}(x)+mM: m=1,2,3,...,
\end{equation}
where M is the Planck mass [B. Zweibach, 2004]. Thus the proposed
event horizon is regarded as enclosing an integral multiple of the
Planck mass.
\par
One class of interactions preserves gravitational equilibrium
outside the horizon by introducing SUSY pairs of mass-less, closed
strings into this domain. The proposed model now focuses upon the
Lagrangians that represent this region and the mass within the
space-like interior of the event horizon. It is argued that this
mass is perceived as an imaginary (tachyonic) mass in the
time-like sector which lies just outside the horizon, and that it
produces vacua that do not preserve that symmetry of the gauge
Lagrangians. If a Lagrangian undergoes a transformation however,
so that it is expanded about the correct vacuum, the tachyonic
mass becomes a real mass, and elements of the adjoint
representation acquire this mass, breaking the gauge symmetry. The
number of gauge bosons that acquire mass is (by the
Nambu-Goldstone theorem) the difference between the number of
elements in the adjoint representation of the broken symmetry and
the number in the adjoint representation of the symmetry which
characterizes the new symmetric domain. With reference to SU(5),
which initially breaks down to SU(3)XSU(2)XU(1), this number is
12. Thus the X and Y particles that mediate the GUT interaction
acquire mass, but the vector bosons of SU(2) do not. Subsequently
however, SU(2) is broken by an analogous process. The hierarchy
between the two symmetry-breaking events is maintained by
renormalization theorems within super-symmetry. In the string
background, the above processes correspond to erosions of the
event horizon, so that strings that were closed and mass-less
again intersect with the branes from which they were temporarily
isolated by the horizon.
\par
The proposed model seems to argue that super-symmetry is never
broken; e.g. that self-realizations of local super-symmetry are
continually re-established by the influx of spin-2 fields into
baryons; i.e. by interactions between spin-2 composites and
valance quarks (quarks experiencing asymptotic freedom), which
manifest themselves as residual SUGRA GUT interactions between
spin-2 fields and spin-(3/2) baryons. And because mass-energy is
preserved, and because symmetry-breaking events reduce the
relevant symmetry to SU(3)XU(1), it is argued that each Planck
mass becomes about $10^{19}$ baryons (1 GeV each). It is
speculated that a resulting fusion of baryons produces a very
dense star and that the resulting inflation event [A. Guth, 1981]
produces the first galaxy.

\section {Quark and Lepton Charges}\label{S:intro}

In the context of the above-described construction, the generators
of flavor SU(3) are regarded as $I_{3}$ and strong hypercharge, Y.
Moreover, Y can be related to $I_{3}$ by the proposed SU(3) flavor
symmetry (so that if $(-1/2)\leq I_{3}\leq (+1/2)$, then
$(-2/3)\leq Y\leq (+2/3)$). In this context, charges, 'Q,' are
assigned to the quarks by
\begin{equation} \label{E:int}
Y=K(\pm I_{3}+Q),
\end{equation}
where K is a proportionality constant to be determined, where in
the absence of Q, Y is proportional (by SU(3) symmetry) to both
+$I_{3}$ and -$I_{3}$ and where the magnitude of the constant of
proportionality is (by symmetry) 1.732. The Y-intercept is Y=KQ.
If the $I_{3}$-Y coordinate system upon which the triplet and
anti-triplet are displayed is centered at the center of symmetry
of two mutually inverted isosceles triangles, then the intercept
occurs at Y=-2/3. Combining this equation with Y=KQ produces
\begin{equation} \label{E:int}
Q=\frac{-2}{3K}.
\end{equation}
In the SU(3) symmetric context moreover, $I_{3}=\pm 1/2$ when
Y=1/3. Thus, one obtains a second equation
\begin{equation} \label{E:int}
\frac{1}{3}=K(\pm \frac{1}{2})+KQ.
\end{equation}
Simultaneous solution of Equations 4.2 and 4.3 yields $K=\pm 2.$
Given the above result, one can describe the isosceles nature of
the triplet by choosing K=+2 for $I_{3}$ less than zero, and K=-2
for $I_{3}$ greater than zero. In this context one can calculate
the charges of the quarks; e.g. Inserting a negative (positive)
slope for a negative (positive) value of $I_{3}$ (dictated by the
slopes of the sides of the isosceles triangle under
consideration), one obtains:
\begin{equation} \label{E:int}
\frac{1}{3}=-2(-\frac{1}{2})+2Q
\end{equation}
and
\begin{equation} \label{E:int}
\frac{1}{3}=+2(+\frac{1}{2})+2Q;
\end{equation}
i.e. one obtains Q=2/3 and Q=-1/3 for the respective values
$I_{3}$=-1/2 and $I_{3}$=+1/2.
\par
The initial state of the postulated universe (determined in the
proposed model by the seven coincident D3-branes that constitute
the initial world volume) involves a chiral doublet and a singlet.
There are two $I_{3}$ options for strings that begin on left
branes and end on color branes. Such strings therefore occur as
doublets:
\begin{equation}
%\[
   \begin{pmatrix}
      U_{L}\\
      D_{L}
   \end{pmatrix}.
%\]
\end{equation}
Similarly there are two $I_{3}$ options for a string that begins
on a left brane and ends on a leptonic brane. Introducing labels
$e^{-}_{L}$ and $\nu^{e^{-}_{L}}$ for such strings, one obtains
doublets such as:
%\[
\begin{equation}
   \begin{pmatrix}
      \nu^{e^{-}}_{L}\\
      e^{-}_{L}
   \end{pmatrix}.
\end{equation}
%\]
On the other hand, there is only one $I_{3}$ option for a string
that begins on a right brane and ends on a leptonic-brane. This is
the option $I_{3}=0$. Describing this asymmetry in terms of what
is called chirality, one states that since $e^{-}_{L}$ and
$e^{-}_{R}$ are of the same charge, both cannot be included within
a chiral system (defined as a system in which opposite helicities
must also be of opposite charge). Thus $e^{-}_{R}$ is an $I_{3}$
singlet, which does not interact chirally (transforming $I_{3}$-up
into $I_{3}$-down or inversely), so that $e^{-}_{R}$ is
characterized by $I_{3}$=0. The same statement applies to $\nu
^{e^{-}}_{R}$. In the same context, $U_{R}$ and $D_{R}$ are
singlets of $I_{3}$=0.
\par
The charges of both quarks and leptons can be determined by weak
hypercharge, but the charges of quarks and leptons are determined
here by flavor symmetry and strong hypercharge because this
approach produces conservation laws in terms of which one can
calibrate the above-described spin-2 adjoint representation of the
symmetry SU(5). The conservation laws $\Delta I_{3}$=0, $\Delta
Y=0$, $\Delta F$=0 and $\Delta$Q=0 permit, and therefore mandate
an interaction
\begin{equation} \label{E:int}
\Sigma^{-}\rightarrow \Lambda^{0}+b+\overline{a}
\end{equation}
(which is confirmed by observation), where $\Sigma ^{-}$ and
$\Lambda$ are baryons, respectively characterized by charges -1
and 0 and by strangeness numbers -1 and -1, where a denotes a
non-baryonic fermion that is characterized by $I_{3}$=+1/2 and b
denotes a non-baryonic fermion of $I_{3}$=-1/2 and where the sum
of the charges of $\overline{a}$ and b is -1. In this context one
can formulate the following three equations:
\begin{equation} \label{E:int}
Q(b)+Q(\overline{a})=-1,
\end{equation}
\begin{equation} \label{E:int}
Q(up)-\Delta Q=a
\end{equation}
and
\begin{equation} \label{E:int}
Q(down)-\Delta Q=b.
\end{equation}
Simultaneous solution yields
\begin{equation} \label{E:int}
\Delta Q=2/3,
\end{equation}
\begin{equation} \label{E:int}
Q(b)=-1
\end{equation}
and
\begin{equation} \label{E:int}
Q(a)=0.
\end{equation}

\section {Large-Scale Structure}\label{S:intro}
The proposed model can be extended to address large-scale
structure. As postulated above, the influx of spin-2 fields into
baryons continually re-establishes a recurring, broken version of
SUSY SU(5), which indicates a recurring inflation event. It will
be demonstrated that six Higgs-plus inflation cycles is equivalent
to an iterative calculation that produces about 3.54X$10^{11}$
galaxies, which is roughly equivalent to the number indicated by
observation. This result depends upon two considerations: one
consists of boundary conditions determined by observations of
local galactic clusters. The other consists of a replacement of
what has traditionally been regarded as a cosmological constant by
the relativistic 4-scaler:
\begin{equation} \label{E:int} \Lambda
=[\frac{h}{\Delta\tau}]^{2}:\Delta \tau=\Delta
t\sqrt{1-\frac{v^{2}}{c^{2}}},
\end{equation}
where h represents Planck's constant and $\tau$ denotes
relativistic proper time. This cosmological parameter is applied
to a model in which each inflation event produces a new group of
4-momentum states that are characterized by a common recessional
speed; i.e. by a new irreducible representation of the Lorentz
group SL(3,1). Elements of each irreducible representation are
distinguished in terms of mass. The first inflation event is
envisioned as producing a galaxy, the second a distribution of
galaxies, the third a distribution of galactic clusters etc. It is
observed that each element of each proposed irreducible
representation corresponds to a Friedman 4-distribution of
mass-energy, and it is postulated that each flat 4-spacetime
distribution is compact (including a boundary beyond which
expansion of the flat distribution does not occur) and simply
connected. Thus, it is postulated that each 4-spacetime
distribution of mass-energy is a 4-hyper-sphere (feasibility
argument is a 4-dimensional analogue of Poincare's conjecture),
enclosing a 5-dimensional space. Finally, the radius of curvature
that is enclosed by each 4-hyper-sphere is regarded as
proportional to the value of the proposed cosmological parameter,
which is characteristic of the distribution (of the hyper-sphere)
in question. Clearly, this radius of curvature is small--very
close to zero--unless the recessional speed of the 4-spacetime
distribution is near the speed of light. For these latter
distributions however, the radii of curvature are significant, so
that the hyper-spherical nature of the distributions is revealed.
According to this model, one should observe, near the edge of the
observable universe, the fragmentation of homogenous shock fronts
into many smaller shock fronts that coincide with outer
hemispheres of the postulated hyper-spheres. This structure is
observed on scales of about $10^{9}$ ly and larger. It was first
observed at Harvard-Smithsonian in 1984 [J. P. Huchra and M.
Geller, 1984]. Let us now consider the iterative calculation that
was described above.

\section {A Theoretical Number of Galaxies}\label{S:intro}

The boundary conditions indicated by observations of the local
galactic clusters are that the number of galaxies in a typical
local cluster is five and that separations of galaxies represent
distances about ten times the galactic diameters; that separations
of clusters represent distances about ten times the diameters of
the clusters etc.
\par
It is determined from observation that the typical galaxy is about
c$t_{0}=10^{5}$ light years (ly) in diameter. The diameter
$ct_{N}$ of the global state that has resulted from the Nth
inflation event is
\begin{equation} \label{E:int}
ct_{N}=ct_{0}10^{N}: N=0,1,2,3,...,
\end{equation}
where the N=0 state corresponds to a single, typical galaxy.
\par
Because five galaxies populate the typical basic cluster, and
because up to a scale of about $10^{9}$ ly, the number of galaxies
can be enlisted as units in terms of which to express volume (up
to this scale galaxies fill a volume rather than populate the
surfaces of semi-spherical shells), one can establish the
following equation to describe the N=1 cluster of galaxies:
\begin{equation} \label{E:int}
\frac{4}{3}\pi R_{1}^{3}=5,
\end{equation}
which implies that
\begin{equation} \label{E:int}
R_{1}=1.06.
\end{equation}
From equation 6.1, the radius of this galactic cluster in ly is
about
\begin{equation} \label{E:int}
ct_{1}=ct_{0}10=10^{5}(10)ly=10^{6}ly.
\end{equation}
One now determines the approximate radius (as a number of
galaxies) of the N=2 state. Given the counting device that is
described by the second of the proposed boundary conditions, it is
argued that the radius of the N=2 state (the separation of the N=1
cluster and a second cluster is ten times the diameter of the N=1
cluster), in terms of a number of galaxies is given by
\begin{equation} \label{E:int}
10(diameter(local cluster))=10(2(1.06))\cong 21.
\end{equation}
Thus, the radius in light years of the N=2 state is (consulting
(6.1))
\begin{equation} \label{E:int}
ct_{2}=ct_{0}10^{2}=10^{5}(10^{2})ly=10^{7}ly.
\end{equation}
Proceeding in this way, one determines the radius (as an
approximate number of galaxies) of the N=3 state, the N=4 state
and the N=5 state. The radius of the N=5 state is (as an
approximate number of galaxies) 168000. Thus, since the radius of
the N=4 state in ly is about $10^{9}$ly, and since (by
observation) galactic clusters populate the surfaces of
semi-spheres on scales larger than $10^{8}$ly, the number of
galaxies in the N=5 state is (summing areas of opposite spherical
shells),
\begin{equation} \label{E:int}
4\pi R_{4}^{2}=4(3.14)(168000)^{2}=3.54X10^{11}.
\end{equation}
The radius in light years of the N=5 state is, consulting equation
6.1,
\begin{equation} \label{E:int}
ct_{5}=ct_{0}10^{5}ly=10^{10}ly,
\end{equation}
which is thought to be the approximate radius of the universe in
terms of Schwarzschild time [J. Towe, 2003, 2006].

%\par
%The problem relating to the cosmological constant can be addressed
%in terms of an argument that what has been treated as a constant
%is actually a relativistic scaler such as $\hbar/\Delta\tau$,
%where $\Delta
%t/\Delta\tau=m/m_{0}=1/\sqrt{1-\frac{v_{\bot}^{2}}{c^{2}}}$, and
%where $v\bot$ denotes the string velocity that is transverse to
%each string. If the kinetic energy that is, in the proposed model,
%radiated from event horizons is characterized as string energy,
%one would be forced to conclude that the radiation of this energy
%produced a very large value of the proposed cosmological
%parameter, while the smaller value of $v\bot$ that associated with
%inflated spacetime returned the parameter to the approximately
%null value with which it is usually associated.
\par

%HERE HERE

%large space volume. This symmetry-breaking event therefore
%corresponds to a major inflation event. Since the postulated, five
%coincident D-branes lie within the event horizon of the black
%hole, unreachable until the black hole has radiated away, the
%strings that represent the pure kinetic energy, which has been
%radiated by the black hole, and now fill 3-space, are decoupled
%from the D-branes, resulting in a closed string representation of
%the electroweak theory:
%\begin{equation} \label{E:int}
%SU(2)_{w}XU(1)_{Y}.
%U(5)\rightarrow U(3)XU(2).
%\end{equation}
%This electroweak symmetry is broken however, when the black hole
%radiates away, so that massless strings intersect with left
%branes. In this context the symmetry
%\begin{equation} \label{E:int}
%U(3)XU(2)=SU(3)_{c}XSU(2)_{w}XU(1)_{Y}XU(1)_{QUAN}
%\end{equation}
%reduces to the standard model
%\begin{equation} \label{E:int}
%SU(3)_{c}XSU(2)_{w}XU(1)_{EM}
%\end{equation}
%coupled with the U(1) symmetry that established the quantization
%conditions which transcended the event horizon.
%\par
%It is now observed that supersymemtry is re-established at the
%thresholds of asymptotic freedom that exist within baryons, and
%that the composite fields that represent the adjoint
%representation of SO(32) mediate interactions with valence quarks
%in domains of asymptotic freedom. It is shown that such
%interactions conserve baryon structure, departing from the SUSY
%GUT tradition that predicts proton decay.

\section{Conclusion}\label{S:intro}
One of the major problems that seems to stand in the way of
physics beyond that standard model is that SUSY GUT theories
predict a proton decay that has not been observed. The foregoing
discussion addresses this by introducing a new SUSY GUT that
preserves baryonic structure. In this model moreover, a broken
version of the SUGRA GUT structure is continually re-established
by the proposed interactions, so that multiple inflation events
are predicted, which address the large-scale structure.
\par
Specifically it was demonstrated that there are $5^{2}-1$ tensor
products of bosons and fermions that are of spin 2; i.e. it was
demonstrated that there is, formally, an adjoint representation of
SU(5) in terms of spin-2 composites provided that one adopts three
fermionic generations and the preservation of generation by the
implicitly indicated GUT interactions. It was then shown that
these composites are physically significant because they are
components of couplings that constitute a locally super-symmetric,
adjoint representation of SU(5).
\par
This SUSY GUT formalism was derived by observing that every quark
can be interpreted as a lepton that is coupled with an appropriate
element of the proposed spin-2 adjoint representation. The same
was noted regarding every lepton. In this context it was
demonstrated that SUGRA interactions between spin-2 composites and
spin-(3/2) baryons can exist as residual manifestations of
quark-lepton transitions within baryonic domains of asymptotic
freedom. Finally it was shown that there are exactly $5^{2}-1$
SUGRA GUT couplings provided that one admits only those
interactions that transform a quark into a lepton or inversely.
The nature of baryons and fermions, as derived from the proposed
Type IIB string theory, motivated association of the preserved
generators of SU(5) with quantum numbers $I_{3}$, strong
hypercharge, fermion number and $\Delta$Q (the charge difference
between a quark and lepton that share the same values of the other
quantum numbers). It was noted that these conservation laws admit
only three types of SUGRA GUT interactions: those that preserve
$I_{3}$=-1/2, $I_{3}$=+1/2 and $I_{3}$=0. These were designated
Types A, B and C and examples were considered. It was demonstrated
that all such interactions preserve baryon structure, departing
from the SUSY GUT tradition that predicts proton decay. It was
noted that all the interactions that were described, plus others
that were implied can also occur within baryons of spin-(1/2),
provided that a photon is absorbed and radiated together with each
spin-2 field---this to preserve local super-symmetry. It was
suggested that the proposed interactions may involve predictions
that permit confirmation of the model. Specifically it was noted
that the proposed interactions indicate that, at any given time,
some triplets within a given baryon may involve three quarks, but
that other triplets may involve replacements including leptons and
spin-2 composites. Thus it was concluded that the proposed model
seems to predict an small-scale inhomogeneity of density within
baryons, and in this context, may lend itself to confirmation.

\par
It was demonstrated that the proposed model can be extended to
address large-scale structure. It was argued that the influx of
spin-2 fields into baryons continually re-establishes a locally
SUSY version of broken SU(5), and that these events produce
recurring inflation events. It was demonstrated that six
Higgs-plus-inflation cycles correspond to an iterative calculation
that produces about $3.54X10^{11}$ galaxies--a number which
corresponds roughly to that indicated by observation.
\par
It was argued that this process and result involve two
considerations. One consists of boundary conditions that are
determined by observations of local clusters. A second involves a
new interpretation of what has traditionally been regarded as the
cosmological constant. It was argued that the cosmological
constant should be replaced by a relativistic 4-scaler:
$\Lambda$=$[\frac{h}{\Delta\tau}]^{2}$. This was motivated by an
intuition that each recurrence of inflation produced a new
irreducible representation of the Lorentz group, which was
characterized by a recessional speed, and the elements of which
were distinguished in terms of mass. It was argued that the first
inflation event produced a distribution of stars, the second a
distribution of galaxies, etc. Finally, because each distribution
was transformed into a flat state by an inflation event, it was
argued that each flat 4-distribution of mass-energy constitutes a
compact 4-space (evidently, galaxies, galactic clusters etc. do
not expand beyond the boundaries that are established at the
conclusions of the inflation events that produce their common
characteristic of flatness). This argument was extended by
postulating that each 4-distribution is simply-connected. Finally
then, it was postulated (parallel to Poincare's conjecture
regarding 3-spaces that are compact and simply connected) that
each such 4-distribution constitutes a 4-spacetime hyper-sphere
that encloses a 5-dimensional space. It was also argued that each
4-hyper-sphere encloses a radius of curvature that is proportional
to the characteristic value of the proposed cosmological 4-scaler
(which is determined by the recessional speed of the 4-momentum
state in question).
\par
Because the value of the cosmological parameter $\Lambda$ is near
zero unless the characteristic recessional speed of a mass-energy
distribution is near the speed of light, it was argued that the
5-dimensional character and the hyper-spherical nature of a
4-momentum state is revealed if and only if that state is receding
at a speed near that of light. Thus it was predicted that shock
fronts near the edge of the observable universe are observed as
fragmented into smaller shock fronts that coincide with the outer
hemispheres of predicted hyper-spheres. It was noted that this
structure is indeed observed at scales of $10^{9}$ ly and larger,
and that it was first observed in 1984 at Harvard-Smithsonian.

\par
$ $\\[-06pt]

%\section{Figures}\label{S:intro}
%\\[15pt]
%\hspace*{64pt}\scalebox{.7}{\includegraphics{figures/Figure
%Eleven.eps}}
%\hspace*{64pt}\scalebox{.7}{\includegraphics{Figure 1.ps}}
%\\[8pt]
%\hspace*{161pt}Figure 1
%\\[15pt]
%It appears that string theory has succeeded in deriving the
%particles that are regarded as constituting the standard model
%from appropriate intersections of strings, D-branes and
%orientifolds [K. Dienes, 1997]. Some argue however that the
%derivation of SM particles from string theory seems contrived, as
%if tailored to obtain the desired result. It is also argued that a
%successful unified theory should provide a feasible SUSY GUT,
%solve the hierarchy problem, account for the breaking of
%electroweak symmetry and for observed inhomogeneities in the large
%scale structure. The discussion to follow will address these
%concerns in terms of an application of Noether's theorem, which is
%applied to local supersymmetry and increments of action that, in a
%context of a preserved symmetry, form the universe.

\end{document}